\documentclass[useAMS]{./mn2e}

\title[Opacity enhancement]{The Solar and $\alpha$ Centauri A and B models improved by opacity enhancement - 
a { possible} explanation for the oversize cool stars
}
\author[M. Y{\i}ld{\i}z]{M. Y{\i}ld{\i}z$^{}$\thanks{E-mail:
mutlu.yildiz@ege.edu.tr},\\
Ege University, Department of Astronomy and Space Sciences, Bornova, 35100
\.Izmir, Turkey}

\input{./psfig}

\begin{document}

\date{Accepted 2006 December 15. Received 2005 December 14; in original form
2006 2005 December 1}

\pagerange{\pageref{firstpage}--\pageref{lastpage}} \pubyear{2006}

\maketitle

\label{firstpage}
\begin{abstract}
{ %
The Sun and $\alpha$ Cen A and B are the nearest stars to us. Despite the general agreement between their
models and seismic and non-seismic constraints, there are serious problems pertaining to their interior.
The good agreement between the sound speed and base radius of the convective zone of the Sun and the solar models 
is broken apart by a recent revision in solar chemical composition. 
For $\alpha$ Cen A and B, however, it is not possible to fit models with the same age and chemical composition to all seismic and non-seismic observational
constraints. 
At the age deduced from seismic constraints, the luminosity ratio ($L_{\rm A}/L_{\rm B}$) of the models
is significantly lower than the ratio taken from the observed luminosities.
Enhancement of opacity as a function of temperature is one way to restore the agreement between solar
models and the Sun,
but such an enhancement does not alter the situation for $\alpha$ Cen A and B. The reason is that 
models of both components are influenced in a similar manner and consequently
the luminosity ratio doesn't change much.
In the present study, problems pertaining to the interior of these three stars with a single
expression for opacity enhancement are modelled.
The opacity enhancement is expressed as a function of density,  ionization degree of heavy elements (oxygen), 
and  temperature.
According to this expression, for improvement of the models the required opacity enhancement for $\alpha$ 
Cen A and B at $\log(T)$= 6.5, for example, is about 7 and 22 per cent, respectively. The enhancement takes 
place in the region in which pressure ionization is effective, and is higher for low-mass stars than for high-mass 
stars. This result seems to be a possible explanation for the serious differences between 
models and observational results of cool stars.  
  
}
\end{abstract}


\section{Introduction}
Interactions of the constituents of our Universe give shape to microscopic and macroscopic objects. In that sense,
opacity, a measure of interaction between electromagnetic waves and particles with mass, is very important for stellar interiors.
The enhancement of opacity to remove discrepancy between models of stellar interioris and seismic and non-seismic constraints on stellar structure 
is an  essential task, because opacity influences both energy production and transfer processes dramatically. 
{ In the present study, such an enhancement is tested for improvement of solar, $\alpha$ Cen A and B models.  
}

Opacity enhancement as a function of temperature has been the subject of many studies.
Simon (1982) considered the influence of opacity enhancement on the period ratio of Cepheid oscillations.
Korzennik \& Ulrich (1989) improved the agreement between theoretical and observed frequencies of oscillation.
Recently, Bahcall, Serenelli \& Pinsonneault  (2004) and Christensen-Dalsgaard et al. (2009) discussed removal of the disagreement
between the sound speeds of the Sun (Basu, Chaplin \& Christensen-Dalsgaard 1997) and current solar models with recently revised solar chemical composition (Asplund, Grevesse \& Sauval 2005,
hereafter, AGS2005).
Zdravkov \& Pamyatnykh considered the effects of opacity enhancement on 
problems pertaining to hybrid stars. The objective of the present paper is to develop a model removing 
the disagreements between interior models of the Sun, $\alpha$ Cen A and B and all their seismic and 
non-seismic constraints.  For these well known stars, { the masses of which are similar but different}, 
the  opacity enhancement is considered as a function of several other physical parameters besides the temperature. 
Such an enhancement is in accordance with 
the results of recent studies on non-ideal effects in the equation of state (EOS: Saumon Chabrier \& van Horn 1995, 
hereafter SCVH).   



{  By using different types of observational techniques, 
the fundamental properties of $\alpha$ Cen A and B are so well known that one can find age of the system from 
these parameters. On the other hand asteroseismology, which is the study of oscillatory 
behavior of stars, puts very important constraints on the internal structure of stars such 
as small and large separations between oscillation frequencies.
The high precision of these constraints for $\alpha$ Cen A and B 
circumscribes the age of the system well.
However, the ages found using the fundamental 
properties and the asteroseismic constraints are rather different:
} 8.9 Gy from the former and 5.5-6
Gy from the latter ({ Eggenberger et al. 2004; Miglio \& Montalban 2005}; Y{\i}ld{\i}z 2007).  At the age derived from the asteroseismic constraints, the ratio of luminosities ($L_{\rm A}/L_{\rm B}$) for models of $\alpha$ Cen A and B
(2.87) is significantly smaller than the ratio from the observed luminosities (3.05). Simple
opacity enhancement as a function of temperature, such as applied in the previous studies, 
affects the structure of $\alpha$ Cen A and B
in a similar manner and cannot solve the problem with the ratio of luminosities. 
However, the physical conditions for a given temperature $T_0$ vary from star to star.
For example, the density at $T_0$ for a low mass star is in general higher than that for a high mass star.
Therefore, the required opacity  enhancement should also depend on density.

Opacity in a medium is a function of cross-section and therefore of the ionization degree
of ions in equilibrium. Equilibrium is the case in which 
ionization and recombination processes balance each other. 
{ Whereas the rate of the ionization process depends upon
the number of energetic photons and henceforth the temperature of the medium, { the density,
which is important for non-ideal effects such as pressure ionization,} 
determines the rate of the recombination process. Thus, ionization equilibrium
is controlled by the density and temperature of the medium.} As a result, opacity enhancement
as a function of temperature is not sufficient. Hence,  we develop a more comprehensive approach for the
enhancement.

For an accurate opacity computation,  the EOS must be computed very precisely. 
In addition to the ionization degree, the excitation  level of an ion is important because 
cross-section of an ion also depends on its excitation level. 
{ In stellar interiors, however, collisional rates for the excitation process 
dominate radiative rates (Mihalas and Weibel-Mihalas 1999). { This is 
another indication showing} 
the importance of density in opacity because the collision rate in a medium
strongly depends on its density.  

For a realistic opacity prediction, atomic data of  
high quality are required. { It is known that although the opacity tables are based on extensive
models and calculation, the calculations do not consider all aspects of the opacity
for a given temperature and density 
(SCVH)
In the present work
we assume that the existing opacity tables are uncertain at certain densities and
temperatures that leave room for modifications.
}
}
%
%

In Y{\i}ld{\i}z et al. (2006), to remove the deviation between model and observed radii (or colors), 
a mixing-length parameter is considered. However, enhancement in opacity,
which controls the energy transfer in the radiative part, increases the size of a star. In addition to that, the opacity in the core 
is also very important for the energy generation rate because the temperature of the core depends on how much released energy 
is permitted to escape.

{
The constraints from oscillation frequencies ($\nu_{n,l}$ with order $n$ and degree $l$) inferred from asteroseismic observations
are { expressed in terms of small ($\nu_{n,l}-\nu_{n-1,l+2}$) 
and large  ($\Delta \nu_{l}(n)=\nu_{n,l}-\nu_{n-1,l}$) separations between these frequencies.
The small separation (Christensen-Dalsgaard 1988) for $l=0$ is defined as
\begin{equation}
\delta \nu_{02}{ (n)}=\nu_{n,0}-\nu_{n-1,2}.
\end{equation}
}
The turning point of oscillations depends mainly on the degree of oscillations $l$. Whereas the oscillations
with $l=0$ sinks down to the center, the turning points of oscillations with $l=1$ and $l=2$ are about $r_{\rm t}=0.05 R_\star$
and $r_{\rm t}=0.10 R_\star$, respectively (Y{\i}ld{\i}z 2008).
As nuclear evolution proceeds, the sound speed throughout the core decreases,
{\ and} the oscillations with $l= 1$ and $l= 2$ sink deeper. The result of this process is { a} reduction in the small 
separation with time.  For the Sun, $\alpha$ Cen A and B,  the oscillations with degree $l=0$, $1$, $2$, $3$ are detected 
(Basu et al. 2007; Bedding et al. 2004; Kjeldsen et al. 2005). { The frequencies of these oscillations
enable } us to compute observational values 
of $\delta \nu_{01}{ (n)}=(\nu_{n0}-(\nu_{n-1,1}+\nu_{n1})/2)$ (Kjeldsen et al. 2005), $\delta \nu_{02}{ (n)}$ and $\delta \nu_{13}{ (n)}$. 

{
The oscillation frequencies are { typically } computed assuming adiabatic process. However, this approximation
is valid for the interior  but not for near-surface regions. 
{ The computation of oscillation frequencies in near-surface regions
requires perturbation of the energy equation. Unfortunately, perturbation 
of the energy equation results in a large uncertainty in the convective flux 
(Christensen-Dalsgaard et al. 1996).}
Kjeldsen Bedding \& Christensen-Dalsgaard (2008) derived an empirical correction for the near-surface 
offset between observed and computed
oscillation frequencies by assuming a power law for the offset. They showed
that this offset is a function of frequency. Use of small and large separations or their ratio (Roxburgh and Vorontsov 2003)
minimizes the effect of the near-surface regions.
}

{ The small separations are a function of $n$.
In the present study, we use their average values over the observed oscillations for each star, for 
comparison with model values. The average values of small ($\delta \nu_{01}$, $\delta \nu_{02}/6$ and 
$\delta \nu_{13}/10$) and large separations ($\Delta \nu_{0}$) for the Sun are calculated 
using the frequencies of an order between $n$=11 and $n$=27. The solar small separations 
are given in Table 1. The { ranges} of $n$ for $\alpha$ Cen A and B are
$n$=16-25 and $n$=20-27, respectively. For l=0, $\delta \nu_{02}/6$ { values of $\alpha$ Cen A and B are} 1.05$\pm$0.12 
and 1.69$\pm$0.08, respectively. { The small separations of some stars are less than zero for some 
values of $n$ 
(see e.g. Soriano \& Vauclair 2008), however, they are always positive for the Sun, $\alpha$ Cen A and B.
}
 
}

}


The remainder of this paper is organized as follows. In Section 2 and 3, we discuss opacity enhancement as a function of temperature in order to improve solar models
and models of $\alpha$ Cen A and B, respectively.
Models with enhancement as a function of several physical parameters including temperature are given in 
Section 4.  { Section 5 is devoted to possible connection between pressure ionization and opacity enhancement.}
Finally, we give
concluding remarks in Section 5.

{


}

\section{Temperature-dependent enhancement to improve solar models}
   \begin{figure}
\centerline{\psfig{figure=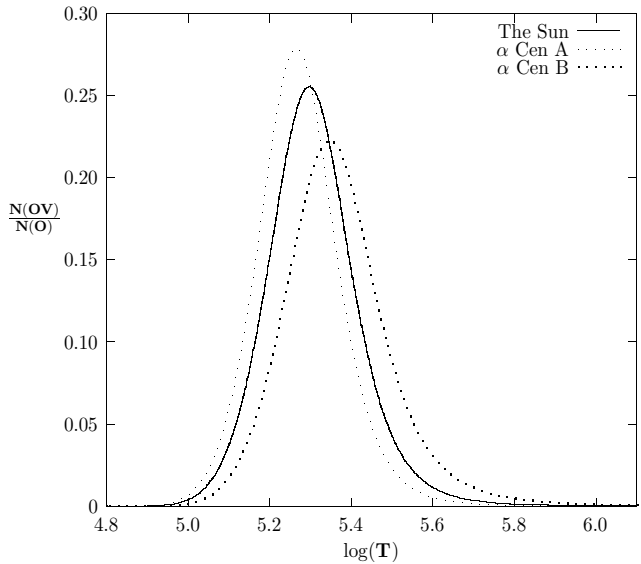,width=170bp,height=213bp}}
      \caption{
Fraction of OV in total number of oxygen ions within the Sun (solid line), $\alpha$ Cen A (thin dotted line) and B (thick dotted line). 
}
              {\label{f1.1}}
   \end{figure}

The degree of ionization has a particular importance in the resistivity of matter against electromagnetic radiation.
In the central regions of stars, the temperature is high enough for almost complete ionization of hydrogen, helium and 
the most abundant heavy elements. Therefore, in these regions the cross-section of ions is very small and the opacity is low.
High ionization occurs in the central regions
of stars while low ionization or neutrality takes place in the outer regions (see Section 5). Specific ions occupy specific parts of the stars.
{ This phenomenon is depicted in Fig. 1.  
The fraction of OV (oxygen ion with four electrons) in the total number of oxygen ions is 
plotted with respect to $\log(T)$ for the models of Sun (solid line) and $\alpha$ Cen A (thin dotted line) 
and B (thick dotted line).} The distribution of OV is likely to follow a Gaussian distribution. 

Therefore, we write the artificial enhancement of opacity as 
\begin{equation}
 \kappa_{\rm enh}=\kappa_{\rm OPAL}+\delta \kappa.
\end{equation}
where $\delta \kappa$ is always positive and represented by a Gaussian function.
\begin{equation}
\delta \kappa=\kappa_{\rm OPAL} c_1~10^{-c_2(log(T)-c_3)^2}.
\end{equation}
In equation (3), $\kappa_{\rm OPAL}$ is the opacity computed from the OPAL tables 
(Iglesias and Rogers 1996) and $c_1$, $c_2$ and $c_3$ are free parameters to 
improve the solar models. { $c_1$ is amplitude of enhancement. { 
While $c_3$ gives temperature ($\log(T)$) at 
which the enhancement is maximum, $c_2$ shows how the enhancement is localized.}}

Solar models are constructed with metallicity $Z=0.016$ for different values of $c_1$, $c_2$ and $c_3$.
The basic properties of the models are listed in Table 1. In the first row, the standard model (Y{\i}ld{\i}z 2008) is given (ModST).
In the second, third and fourth columns, the numerical values of the free parameters $c_1$, $c_2$ and $c_3$ are listed, respectively. 
For $c_1=0.25$, the solar models for three different values of $c_2$ are given in the second (ModK1), third  (ModK2) and fourth (ModK3) rows of 
Table 1.
A decrease in $c_2$ reduces the base radius of the solar convective zone ($r_{\rm bcz}$), however it augments the helium abundance in the solar envelope ($Y_{\rm s}$).
The fifth row gives the ModK4 solar model. For this model, opacity { enhancement is  
more localized to} the regions just below the convective zone. Its $Y_{\rm s}$ and $r_{\rm bcz}$ are in very good agreement 
with the observed values.
In addition to the good agreement between the observed and predicted base radius of convective zone, 
the small separations { ($\delta \nu_{01}$,  $\delta \nu_{02}$ and  $\delta \nu_{13}$)} and the sound speed concur reasonably well. 
The relative sound-speed differences between these models and the Sun (RSSD)
are plotted in  Fig. 2. The RSSD for the standard solar model (ModST)  is about 1.7 per cent and decreases with enhanced opacity below the convective zone. 
Thin dotted, thick solid, and thick dotted lines in Fig. 2 show ModK1, ModK2, and ModK3, correspondingly. Among these, ModK2 ($c_2=4.5$) is in better agreement with the Sun than the others.
Another solar model (ModK4) in which enhancement of $\kappa$ is
{ more localized to 
below the} convective zone ($c_1=0.30$ and $c_2=20$) gives better results for $Y_{\rm s}$ and $r_{\rm bcz}$ than the solar models with $c_1=0.25$.

A better model (ModKM in the { sixth} row of Table 1) { is} constructed with the 
mean of enhancements of ModK2 and ModK4. This ModKM model is the best of the solar models
regarding all seismic constraints and including the small separations given in the last three columns of Table 1. 
The RSSD for the ModKM model is less than 0.2 per cent (solid line with diamonds in Fig. 2).
Also shown in Fig. 2 is the RSSD for the solar model constructed with opacity enhancement (dotted line with $+$) as given in equation (4)  of Bahcall et al. (2004).
The maximum RSSD for the Bahcall et al. (2004) model is about 0.6\%. Its basic properties are given in Table 1 (ModB4).


\begin{table*}
\begin{center}
      \caption{ Basic properties of the solar models with the chemical composition given by AGS2005.
The last line is for the observed values.
The uncertainty in $R_c$ is 0.001 $R_\odot$ (Basu \& Antia 1997).
The observed values of the surface helium and metal abundances are taken from Basu \& Antia (1995) and AGS2005, respectively.
The unit of small separations $\delta \nu_{01}$, $\delta \nu_{02}$, and $\delta \nu_{13}$ (Basu et al. 2007) is $\mu$Hz.
}
{\label{t1.1}}
{\begin{tabular}{lcrccccccccclllll}
\hline
Model&$c_1$&$c_2$~&$c_3$&$X_0$  &$\alpha$&$\rho_c$& $T_c$ &$X_c$ &$ Z_s $  &~~ $Y_s$& $R_c/R_\odot$ & $\delta \nu_{01}$& $\delta \nu_{02}/6$& $\delta \nu_{13}/10$ \\  
\hline
ModST&---  & ---  & --- & 0.70975 &  1.820  & 146.78 & 15.65 & 0.355 & 0.0124 &0.244&0.733  & 1.879 & 1.686 &1.739\\  
ModK1&0.25 &18.0  & 6.30& 0.70901 &  1.871  & 146.85 & 15.65 & 0.355 & 0.0128 &0.248&0.7162 & 1.821 & 1.662 &1.716 \\  
ModK2&0.25 & 4.5  & 6.30& 0.70506 &  1.890  & 148.46 & 15.70 & 0.349 & 0.0129 &0.252&0.7130 & 1.811 & 1.646 &1.701 \\  
ModK3&0.25 & 3.0  & 6.30& 0.70200 &  1.906  & 149.35 & 15.72 & 0.346 & 0.0129 &0.255&0.7125 & 1.800 & 1.638 &1.696 \\  
ModK4&0.30 &20.0  & 6.32& 0.70881 &  1.877  & 146.93 & 15.65 & 0.354 & 0.0128 &0.248&0.7130 & 1.821 & 1.661 &1.715\\  
ModKM&---  & ---  & --- & 0.70684 &  1.883  & 147.76 & 15.68 & 0.352 & 0.0128 &0.250&0.7130 & 1.814 & 1.653 &1.709\\  
ModB4&---   & --- & --- & 0.70898 &  1.857  & 146.80 & 15.65 & 0.355 & 0.0128 &0.248&0.7140 & 1.811 & 1.688 &1.740\\  
Obs &---  & ---   & --- & --- & ---   & ---  & ---  & ---  & 0.0122 &0.25&0.713            & 1.813 & 1.653 &1.683 \\  

\hline
\end{tabular}}
\end{center}
\end{table*}
   \begin{figure}
\centerline{\psfig{figure=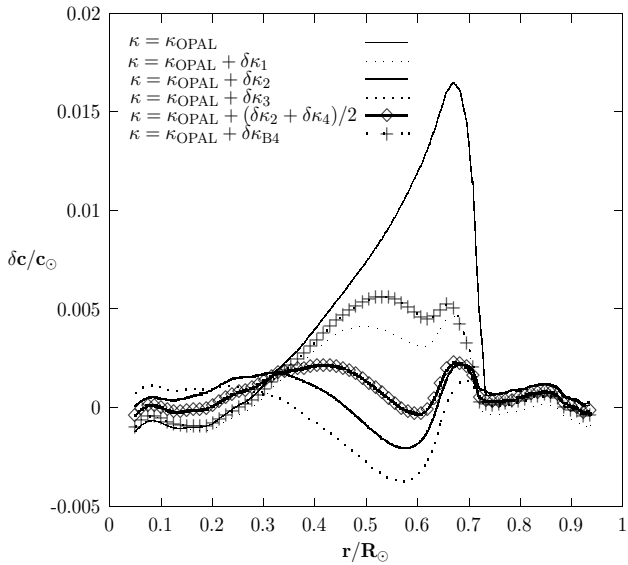,width=170bp,height=213bp}}
      \caption{
The relative sound-speed difference between the solar models given in the first three rows of Table 1 and the Sun is plotted with respect to the relative
                radius. The thin solid line shows the standard solar model with $Z_0=0.016$ (Y{\i}ld{\i}z 2008).
The other models are with enhanced opacity and given in Table 1 (see the text).
}
              {\label{f1.1}}
   \end{figure}
%


\section{Models of $\alpha$ Cen A and B with enhanced  opacity as a function of temperature}
\begin{table*}
\label{ta2}
\caption{
 Model properties of $\alpha$ Cen A and B. $ \delta \nu_{02}$ and $ \Delta \nu_0$ are  in units $\mu$Hz.
The error in $ \Delta \nu_0$ is 0.1 $\mu$Hz. The uncertainty of
$ \delta \nu_{02} $ is 0.12 $\mu$Hz for $\alpha$ Cen A and 0.08 $\mu$Hz for $\alpha$ Cen B.  }
$
\begin{array}{llllcclllll}
\hline 
\noalign{\smallskip}
Star&L_{\rm }/L_\odot& R_{\rm }/R_\odot & T_{\rm eff}& X_{\rm 0}& Z_{\rm 0} & \alpha & t(10^9 y) & \delta \nu_{02}/6& \Delta \nu_0& MODEL\\
\noalign{\smallskip}
\hline
\noalign{\smallskip}
A  & 1.549& 1.223& 5828& 0.7129& 0.023&  1.710 & 5.90& 0.91&106.7& SIS23\\
B  & 0.569& 0.863& 5400& 0.7129& 0.023&  1.980 & 5.90& 1.70&164.4& SIS23\\
\hline
B  & 0.507& 0.864& 5246& 0.7129& 0.023&  1.820 & 5.90& 1.59&164.7& ModK3\\     
\hline
{ A}  & 1.535& 1.224& 5824& 0.713& 0.023&  1.71 & 5.90& 0.97&106.4& SIS23KBah\\
{ B}  & 0.567& 0.863& 5395& 0.713& 0.023&  1.98 & 5.90& 1.56&164.5& SIS23KBah\\
\hline
A  & 1.545& 1.226& 5817& 0.7066& 0.023&  1.881 & 7.20& 0.65&106.4& ModK5z23\\ 
B  & 0.506& 0.862& 5247& 0.7066& 0.023&  1.930 & 7.20& 1.45&165.0& ModK5z23\\          
\hline
A  & 1.546& 1.226& 5817& 0.7066& 0.023&  1.731 & 5.90& 0.97&106.1& ModK6z23\\ 
B  & 0.507& 0.863& 5249& 0.7066& 0.023&  1.791 & 5.90& 1.60&165.0& ModK6z23\\          
\hline
A  & 1.545& 1.223& 5818& 0.7135& 0.021&  1.676 & 5.56& 1.04&106.7& ModK6z21\\                
B  & 0.506& 0.863& 5246& 0.7135& 0.021&  1.775 & 5.56& 1.64&164.9& ModK6z21\\
\hline
A  & 1.543& 1.224& 5819& 0.7126& 0.019&  1.743 & 5.58& 1.01&106.7& ModK8z19\\                
B  & 0.507& 0.863& 5248& 0.7126& 0.019&  1.874 & 5.58& 1.62&164.9& ModK8z19\\
\hline
A  & 1.544& 1.224& 5824& .... & .....&  ... & .... & 1.05&105.5& { obs}\\
B  & 0.507& 0.863& 5250& .... & .....&  ... & .... & 1.69&161.5& { obs}\\
\hline
\noalign{\smallskip}
\hline
\end{array}
$
\end{table*}
\begin{table*}
\begin{center}
      \caption{ 
Basic physical conditions at the bottom of the convective zones of the standard solar and $\alpha$ Cen A and B models and amount of opacity enhancement there
to improve these models.
}
{\label{t1.1}}
{\begin{tabular}{lcccc}
\hline
Star~~  &  $M_c/M_\odot$ & $T_{\rm bcz}$(MK)  & $\rho_{\rm bcz}$(g~cm$^{-3}$) & $c_1$ \\
\hline
$\alpha$ Cen B  & 0.934   & 2.69  & 0.571 & 0.425\\
The Sun & 1.000   & 1.97  & 0.150 & 0.250\\
$\alpha$ Cen A  & 1.105   & 1.73  & 0.068 & 0\\

\hline
\end{tabular}}
\end{center}
\end{table*}

\begin{table*}
\begin{center}
      \caption{ Basic properties of the solar models with enhanced opacity as a function of $\rho$, $T$, and ionization degree of oxygen.
See also Table 1.
}
{\label{t1.1}}
{\begin{tabular}{lcccccccclllll}
\hline
Model&$X_0$  &$\alpha$&$Z_0$& $T_c$ &$X_c$ &$ Z_s $  &~~ $Y_s$& $R_c/R_\odot$ & $\delta \nu_{01}$& $\delta \nu_{02}/6$& $\delta \nu_{13}/10$ \\  
\hline
ModK5& 0.70174 &  1.875  & 0.0160 & 15.71 & 0.349 & 0.0128 &0.255&0.7167 & 1.801  & 1.635  &1.691 \\  
ModK6& 0.70163 &  1.885  & 0.0153 & 15.67 & 0.353 & 0.0124 &0.257&0.7111 & 1.811  & 1.688  &1.740\\  
ModK7& 0.70106 &  1.898  & 0.0155 & 15.63 & 0.357 & 0.0125 &0.248&0.7130 & 1.811  & 1.660  &1.716 \\  
ModK8& 0.70854 &  1.892  & 0.0150 & 15.61 & 0.357 & 0.0121 &0.250&0.7130 & 1.809  & 1.655  &1.710 \\  
Obs & --- & ---   & ---  & ---  & ---  & 0.0122 &0.25&0.713            & 1.813 & 1.653 &1.683 \\  

\hline
\end{tabular}}
\end{center}
\end{table*}

Treating the enhanced opacity as a function of temperature improves the solar models. If we apply 
any of the enhancements discussed in the previous section, for example that of Model 3, 
to models of $\alpha$ Cen A and B with chemical composition given by Feltzing \& Gonzalez (2001), 
{
the fundamental properties ($L$, $R$, $T_{\rm eff}$) of the models of $\alpha$ Cen A and B are essentially unchanged.
The same is true for the opacity enhanced as in Bahcall et al. (2004). 
The basic properties of models of $\alpha$ Cen A and B with the enhanced opacity of Bahcall et al. 
(2004)
are listed in the fourth and fifth rows of Table 2 (SIS23KBah), respectively. 
These models are obtained with the parameters of Model SIS23 of Y{\i}ld{\i}z (2008). 
For comparison, the basic properties of Model SIS23  are also listed in the first and second rows of Table 2. 
Despite the negligibly small differences between the luminosities and radii 
of the models with and without opacity enhancement, the small separations of Models SIS23KBah for $\alpha$ Cen A and B (with a moderate
enhancement of opacity) are slightly
different from those of Models SIS23. This is consistent with the result of previous section that the opacity enhancement 
influences the sound-speed profile, at least in the central regions.   
However, the effects  on $\alpha$ Cen A and B are in opposite directions; 
the enhancement decreases $\delta \nu_{\rm 02}$ of $\alpha$ Cen B but increases $\delta \nu_{\rm 02}$ of $\alpha$ Cen A.
Therefore, { the temperature-dependent opacity enhancement concept does not improve the $\alpha$ Cen A and B models}.


}

\section{The generalized enhancement of opacity }

{ The agreement between observed and model predicted luminosity values of $\alpha$ Cen B may be improved by 
increasing the coefficient $c_1$. For this purpose, we set $c_1$=0.425. The results from the $c_1$=0.425 model are 
 given in the third row of Table 2.  }
Since  model SIS23 of $\alpha$ Cen A is already in very good agreement with seismic and non-seismic constraints, opacity enhancement is not { needed. Hence, the opacity enhancement factor} is assumed to be zero for $\alpha$ Cen A. Next, the $c_1$ factor is considered to be a function of stellar mass. 
The basic physical conditions at the bottom of the convective zones of the standard solar and $\alpha$ Cen A and B models are listed 
in Table 3. The second column in Table 3 lists for the masses of the stars. The third and the fourth columns report
the base temperature and density of the convective zones. In the fifth column, the values of coefficient $c_1$,
nearly the maximum value of $\delta \kappa / \kappa$,
are given.
Despite the very different 
chemical composition of the models for the Sun and the components of $\alpha$ Cen, 
the relation between  the factor $c_1$ and stellar mass is nearly linear (see Fig. 3). 
This relation can be attributed to density dependence of $c_1$. The densities below the 
convective zones of the Sun, $\alpha$ Cen A and B are very different from each other. For example,
the density at the base of the convective zone of $\alpha$ Cen B is about four and eight times higher than those of the
Sun and $\alpha$ Cen A, respectively. 
{ Since the base temperatures of the convective zones of these stars are much closer to each other than the densities, $c_1$ } can  be considered as a function of density only.
The density ranges of the regions in which opacity is effectively enhanced are very different for the Sun and $\alpha$ Cen A and B, and 
move toward higher densities for low mass stars.

   \begin{figure}
\centerline{\psfig{figure=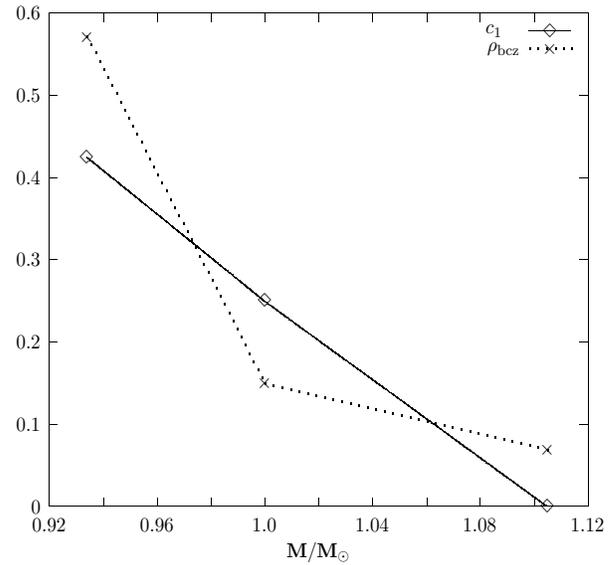,width=170bp,height=213bp}}
      \caption{
$c_1$ and base density of convective zone as a function of stellar mass.
}
              {\label{f1.1}}
   \end{figure}

   \begin{figure}
\centerline{\psfig{figure=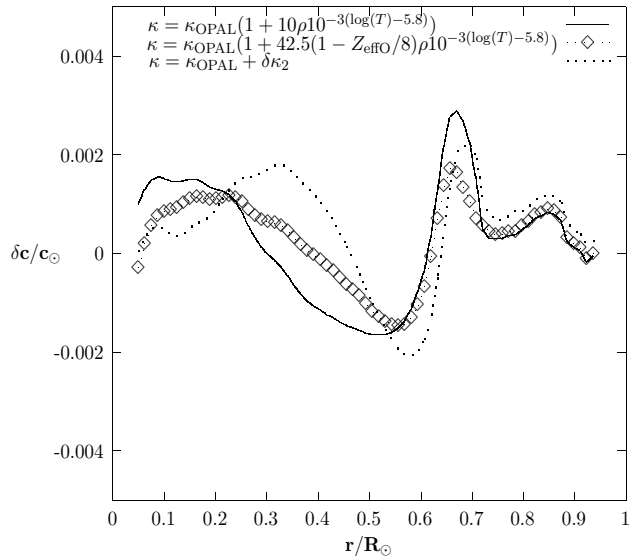,width=170bp,height=213bp}}
      \caption{
The relative sound-speed difference between the solar models and the Sun.
}
              {\label{f1.1}}
   \end{figure}

   \begin{figure}
\centerline{\psfig{figure=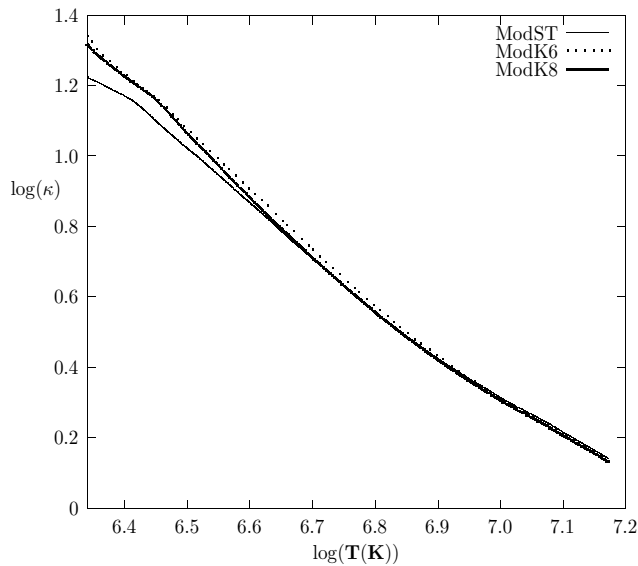,width=170bp,height=213bp}}
      \caption{
The enhanced opacity in the Sun with respect to log(T). The base temperature of the convective zone ($T_{\rm bcz}$) of the Sun is given by $\log T_{\rm bcz}$=6.34. 
The maximum enhancement ($\delta\kappa/\kappa$) for the solar model is about 0.1.
}
              {\label{f1.1}}
   \end{figure}
   \begin{figure}
\centerline{\psfig{figure=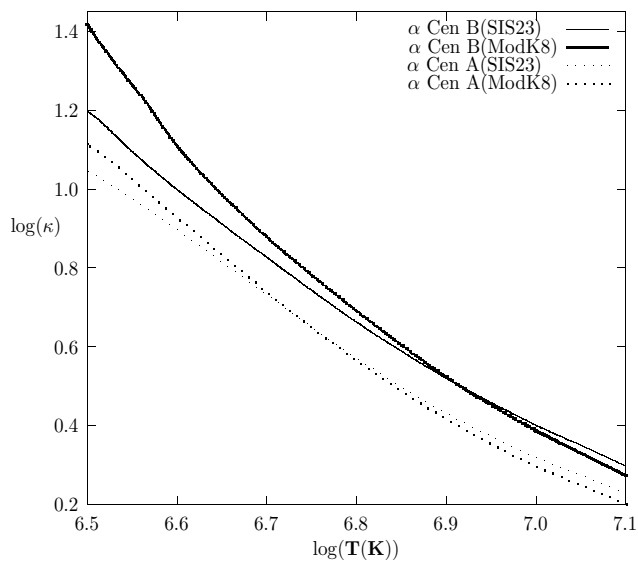,width=170bp,height=213bp}}
      \caption{
The enhanced opacity in $\alpha$ Cen A and B with respect to $\log(T)$. The base temperatures of the convective zones ($T_{\rm bcz}$) of $\alpha$ Cen A and B are different: $\log T_{\rm bcz}$ 
of $\alpha$ Cen A and B are 6.31 and 6.50, respectively. The enhancement ($\delta\kappa/\kappa$) at $\log T_{\rm bcz}$=6.5  is about 0.067 for model of $\alpha$ Cen A and 0.22 for model of $\alpha$ Cen B.
}
              {\label{f1.1}}
   \end{figure}

As stated above, the dependence of $c_1$ on stellar mass can be related to density in the region
below the convective zone, which differs from star to star.
Therefore, some expressions for opacity enhancement as a function of density
and temperature, which improve the solar and the $\alpha$ Cen A and B models, can be tested.
{ For this purpose}, $c_1$ is plotted as a function of { stellar} mass in Fig. 3 (solid line with diamonds).
Also plotted in Fig. 3 is the base density of the convective zones ($\rho_{\rm bcz}$) for each star. There is a 
correlation between $\rho_{\rm bcz}$ and $c_1$. We therefore construct a solar model with 
$\delta\kappa_\rho= 10\rho 10^{-3(\log T - 5.8)} $. The characteristics of this solar model are given in the first row of Table 4 (ModK5).
{ The seismic properties from the new model are not as good as those of the solar models with temperature-dependent 
opacity enhancement. However, they are
much better than those of the standard model ModST.}
The small RSSD for ModK5 is plotted with respect to relative radius in Fig. 4 (solid line). 

{
We also construct models for $\alpha$ Cen A and B with $\delta\kappa_\rho$. The basic properties of these models 
 (ModK5z23) are listed in Table 2.
The age from the fundamental properties of $\alpha$ Cen A and B is { about} 7.2 Gyr. At this { estimated age},
the model values of small separations are significantly smaller than the observed values. 
Despite the improvement,
the ages derived from the fundamental properties and the seismic constraints are not in good agreement;
the age of the models is about 1.5 Gyr greater than the seismic age.
}

$\delta\kappa_\rho$ given above may solve the { inconsistencies } pertaining to the solar interior; 
{ however,
it does not help with $\alpha$ Cen A and B}. { In order to obtain good agreement between the ages of $\alpha$ Cen A and B from fundamental properties and asteroseismic properties, the coefficient of $\delta\kappa_\rho$ should be increased. 
{ Unfortunately, such a treatment hurts the agreement between the solar model and the Sun}. 
Therefore, further parameters are required for opacity enhancement.}

{ 
The problem with $c_1$ as a linear function of just $\rho$ is that it increases toward
the center of stars and becomes maximum at the center. On the other hand, at the
center of stars the opacity is essentially very low as a result of high
ionization and therefore no or little enhancement is expected. For the radiative 
interior of cool stars, the effective charge (mean ionization degree) of oxygen 
(${Z_{\rm eff,O}}$) is a good indicator for opacity.  For a completely ionized medium, 
where ${Z_{\rm eff,O}}$ is equal to atomic number of oxygen ($Z_{\rm O}$), a very low 
value of opacity is determined by free-free transitions. In that case, we assume no 
opacity enhancement. Therefore, to minimize opacity enhancement near the 
center, we multiply $\delta\kappa_\rho$ by 
$(1-\frac {Z_{\rm eff,O}} {Z_{\rm O}})^{1/2}$, which is equal to zero for the
completely ionized medium (see below).
}

The EOS routines of the code that we use solve the Saha equation (Y{\i}ld{\i}z and K{\i}z{\i}lo\u{g}lu 1997). 
{ The Debye- H\"uckel approximation is employed for Coulomb effect and EOS of Mihalas et al. (1990, hereafter MHD) 
is used to compute the partition functions.
The ionization degree of the ten most abundant elements is computed accurately.} Oxygen
is the most abundant heavy element in the Sun and therefore opacity is very 
sensitive to the degree of its ionization, at least in the inner radiative region. 
Therefore, we enhance the OPAL opacity as 
\begin{equation}
\kappa_{\rm enh}=\kappa_{\rm OPAL} (1+42.5\rho (1-\frac {Z_{\rm eff,O}} {Z_{\rm O}})^{1/2}~10^{-3(\log(T)-5.8)^2}).
\end{equation}
{
In terms of the number of oxygen ions with charge $i$ ($N_i$), 
${Z_{\rm eff,O}}=\frac{1}{N_{\rm O}}\sum i N_i $, where $N_{\rm O}$ is the total number of oxygen ions.}

When it comes to sound speed, the solar model with the latest opacity enhancement { (ModK6)} agrees the best among all the solar models. The results from the ModK6 model are shown in Fig. 4. The 
dotted line with diamonds in Fig. 4 represents this model. The maximum value of the relative difference is about 0.15 per cent.
The basic properties of the solar model ModK6 are given in the second row of Table 4. Its $Y_{\rm s}$, $\delta \nu_{\rm 02}$ and $\delta \nu_{\rm 13}$ are slightly greater
than the observed values given in the fifth row of Table 4. 

We also construct  models for $\alpha$ Cen A and B with the enhanced opacity given in equation (4) for Z=0.023 (ModK6z23) and 0.021 (ModK6z21). The properties of these
models are reported in Table 2. The small separations ($\delta \nu_{\rm 02}$) of both models are in agreement with the asteroseismic inferences.
However, the models with Z=0.021 display better agreement than those with Z=0.023.

We also obtain solar models based on modified version of  equation (4): 
\begin{equation}
\kappa_{\rm enh}=\kappa_{\rm OPAL} (1+4.5\rho (1-\frac {Z_{\rm eff,O}} {Z_{\rm O}})^{1/2}~10^{-17(\log(T)-6.32)^2}).
\end{equation}
The solar model with this opacity (ModK7) is given in the third row of Table 4. Its $Y_{\rm s}$, $\delta \nu_{\rm 02}$ and $\delta \nu_{\rm 13}$ are in better agreement 
with the observed values than that of ModK6. 

Also given in Table 4 is ModK8, the opacity of which is taken as the mean of equations (4) and (5). This model is also in very good agreement with the Sun.
The models of $\alpha$ Cen A and B with this opacity and Z= 0.019 are also constructed; ModK8z19 A and ModK8z19 B in Table 2 are also in good agreement with their
seismic and non-seismic constraints.

{
The opacity enhancement of ModK5s (ModK5 for the Sun and ModK5z23 for $\alpha$ Cen A and B) is a function of $\rho$ and $T$; 
$f(\rho,T)=\delta\kappa_\rho= 10\rho 10^{-3(\log T - 5.8)} $. For ModK6s (ModK6 for the Sun,  ModK6z23 and ModK6z21 for $\alpha$ Cen A and B) and ModK7, the enhancement
is as given in equations (4)-(5), respectively. 
The main difference between these enhancements is that equations (4)-(5) are also a function of 
ionization degree (effective charge) of heavy elements (oxygen). Opacity in the (radiative) interiors of the Sun, $\alpha$ Cen A and B
is essentially determined by bound-free transitions (ionization). However, the effect of this process is null if 
the gas is completely ionized. In this case, $\frac {Z_{\rm eff,O}} {Z_{\rm O}}$ becomes unity.
Then, no enhancement is required  when the term $(1-\frac {Z_{\rm eff,O}} {Z_{\rm O}})$ in equations (4)-(5) is zero.  
Since the models with enhancements with equations (4)-(5) are in good agreement with the observed properties of 
these stars, we deduce that i) simple enhancement as a function of $\rho$ and $T$ is inadequate, and ii)
much more precise methods for the computation of ionization and cross-sections of heavy elements are required 
than used at present.
}

{
\subsection{Comparison of opacities of standard and improved models for the Sun, $\alpha$ Cen A and B}

The opacities ($\log \kappa$) of standard and improved solar models are plotted with respect to $\log (T)$ in Fig. 5. 
The thin solid line shows the standard model (ModST), which has the lowest opacity in the outer part of the 
solar radiative core.  The thin dotted line and thick solid line represent  
the improved solar models ModK6 and ModK8, respectively. In the core, the opacities of solar models are 
very close to each other; in other words, the enhancement is not needed there. In the outer parts of the solar 
radiative core, however, the opacities of improved models are significantly higher than that of the standard model. 
The opacity difference between the standard model and 
ModK8 is remarkable for $\log (T)< 6.6$. The opacity of  ModK6, however, is the same as that of the standard 
model only in the central regions ($\log (T)> 6.9$).

In Fig. 6, we compare opacities of standard (SIS23, thin lines) and improved (ModK8z19, thick lines)  models for 
$\alpha$ Cen A and B. The opacities ($\log \kappa$) in the interior of $\alpha$ Cen A (dotted lines) and B (solid lines) are 
plotted with respect to $\log (T)$.
It should be noticed that the chemical composition of standard and improved models are not the same and therefore 
the difference between the opacities is not entirely due to the enhancement. Due to a lower metallicity of improved 
models than the standard models, despite the 
enhancement 
the opacities of improved models near the central regions of $\alpha$ Cen A and B are lower than those of their standard counterparts.
In the outer part of inner radiative regions of $\alpha$ Cen A and B, the enhancement is effective for $\log (T) < 6.7$ 
and $\log (T) = 6.9$, respectively. 
The opacity difference between the improved and standard models of $\alpha$ Cen B is greater than the difference for 
$\alpha$ Cen A. This result is due to the density and effective charge dependence of the enhancement applied to the improved 
models.
}

The bases of the solar and $\alpha$ Cen B models have very different densities  and temperatures ($T_{\rm bcz}$).
$\log T_{\rm bcz}=6.34$ for the Sun and $\log T_{\rm bcz}=6.5$ for $\alpha$ Cen B. The enhancement 
($\delta\kappa/\kappa$) at the base of the convective zones
needed to improve the solar and $\alpha$ Cen B models  is about 0.1 and 0.17, respectively. 

{ 
For assessment of the density dependence of opacity enhancement, it is better to give the amount of enhancement at the same temperature.
For example, at $T=2.69 \times 10^6 $ K, the base temperature of convective zone of  $\alpha$ Cen B, 
the amount of opacity enhancement ($\delta \kappa/\kappa$) is 0.067 for $\alpha$ Cen A and 0.22 for $\alpha$ Cen B (ModK8z19 with $Z_0=0.019$).
For the improvement of the solar model (ModK8 with $Z_0=0.015$), the amount of enhancement at the same temperature is 0.042.

\subsection{Comparison of evolutionary tracks of standard and improved models for $\alpha$ Cen A and B}
   \begin{figure}
\centerline{\psfig{figure=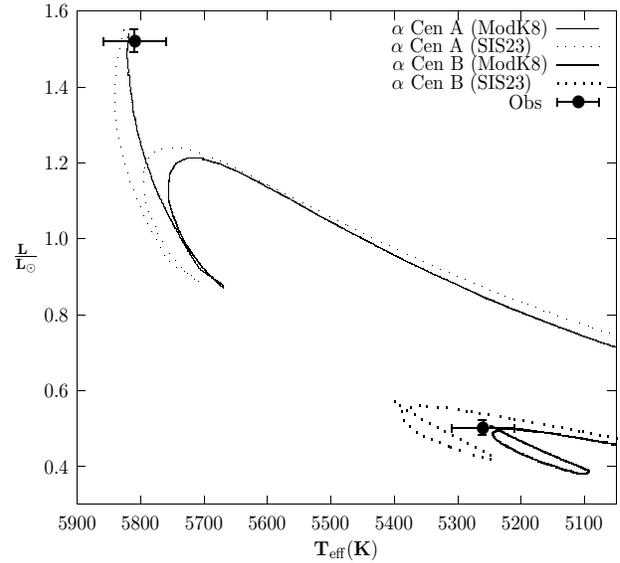,width=170bp,height=213bp}}
      \caption{
Evolutionary tracks of models for $\alpha$ Cen A and B in the HR diagram. While the dotted lines show 
the standard models  (SIS23) of $\alpha$ Cen A (thin) and B (thick), the solid lines represent 
the improved models (ModK8) for $\alpha$ Cen A (thin) and B (thick).
}
              {\label{f1.1}}
   \end{figure}
Changes in interior opacity significantly affect stellar structure as a whole.
The opacity enhancement given above alters evolutionary tracks of cool stars, in particular.
In Fig. 7, the standard and improved models are plotted in a theoretical HR diagram. 
The observed positions of $\alpha$ Cen A and B (circle) are also seen. 
The dotted lines show the standard models while the improved models are represented by the solid lines.
The evolutionary tracks comprise main-sequence (MS) and pre-MS stages.
Due to the density and effective charge dependence of the opacity enhancement for ModK8, 
during the MS stage
the tracks for $\alpha$ Cen A are much closer to each other than those of $\alpha$ Cen B.
For $\alpha$ Cen A, both the standard and improved models are in good agreement with the 
observed results. However, the tracks near the zero-age MS (ZAMS) phase and in the greater part of the MS phase 
are slightly different.
The effective temperature difference between the tracks during the MS (except near the terminal-age MS) 
phase is about 40 K.
For $\alpha$ Cen B, however, the difference is about 150 K. The improved model of $\alpha$ Cen B is in good agreement
with the observed position. The luminosity difference between the ZAMS of the tracks is about 0.04 $L_\odot$ 
and reaches 0.06 $L_\odot$ for the present models. 
For a given value of luminosity, the radius of the improved model of $\alpha$ Cen B is 4 per cent greater than that of 
the standard model. 
{ In the following subsection, we consider stellar models with mass lower than that of $\alpha$ Cen B.
} 
%


%
}
{ 
\subsection{Influence of opacity enhancement in structure of 0.8 and 0.9 M$_\odot$ models}
   \begin{figure}
\centerline{\psfig{figure=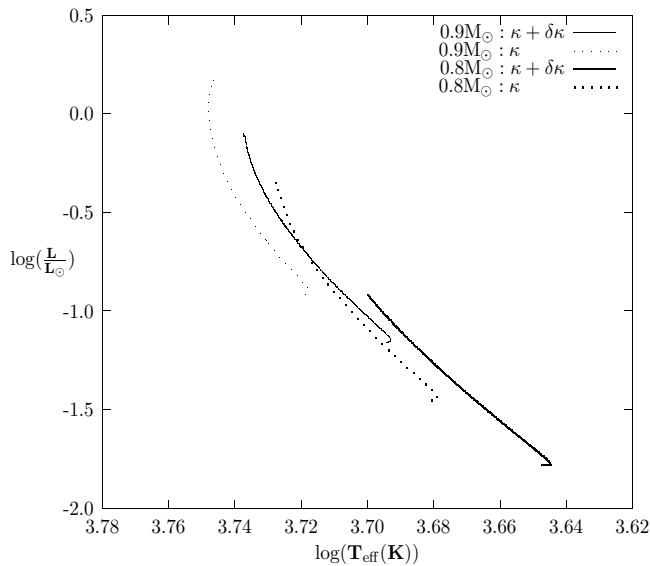,width=170bp,height=213bp}}
      \caption{
Evolutionary tracks of 0.8-(thick) and 0.9-M$_\odot$ (thin) models with ModK8z19 values of $\alpha$ Cen B. 
While the dotted lines show
the standard models, the solid lines represent
the models with enhanced opacity.
}
              {\label{f1.1}}
   \end{figure}

In order to test influence of the opacity enhancement on the structure of stars 
cooler than $\alpha$ Cen B, 
we construct 0.8 and 0.9 M$_\odot$ models with ModK8z19 values of $\alpha$ Cen B:
$X=0.7126$, $Z=0.019$  and $\alpha=1.874$. In Fig. 8, the evolutionary tracks of 
these models with enhanced  (solid lines) and standard (dotted line) opacities are plotted in the HR diagram. 
Opacity enhancement seriously decreases the ZAMS luminosity; $\delta \log L=\log(L_{\rm enh}/L_{\rm st})$ is -0.14 for 0.8M$_\odot$ and -0.11  for 
0.9M$_\odot$, where enh and st represent improved and standard models, respectively. 
Radii of models for a given mass is almost the same near the ZAMS. However, the effective 
temperature difference between the enhanced and the standard models ($T_{\rm eff,enh}-T_{\rm eff,st}$) 
is -357 K for  0.8 M$_\odot$ and -286 K for 0.9 M$_\odot$.

We can also compare sizes of standard and enhanced models of the same luminosity.
For $L=$0.25L$_\odot$, for example, the difference between the radii of standard and enhanced models 
of 0.8M$_\odot$ is about 4.6 per cent. The same amount of difference also exists for the 0.9 M$_\odot$ model for L=0.41 and 0.90L$_\odot$. The temperature difference is now about 100 K for both models for any fixed value of luminosity.
As a result, opacity enhancement may be a possible explanation for why low-mass stars are cooler than (or oversize compared with) 
their interior models (see  e.g. Clausen et al. 2009).

}
{ 
\section{Pressure ionization and opacity enhancement}
It is very difficult to reveal the physics underlying opacity enhancement. 
However, it seems that the major role is played by the non-ideal effects of pressure (density) ionization. 

Our EOS takes pressure ionization into account by using the classical Debye-H\"uckel approximation (DHA) 
(Y{\i}ld{\i}z and K{\i}z{\i}lo\u{g}lu 1997).
Detailed analysis of Coulomb interactions between ion-ion, ion-electron and electron-electron 
shows that the DHA overestimates the Coulomb effect. 
SCVH, for example, consider the ionization 
of hydrogen at low temperatures and increasing densities via a first-order 
'plasma phase transition' 
and find the existence of a significant amount of neutral hydrogen  (5 per cent) at the bottom of the convective zone, 
where the DHA predicts complete ionization (see fig. 24 of SCVH). Such a difference is not very important for the pressure
(fig.25 of SCVH); the maximum difference between pressures of the two EOS is about 1 per cent.  
For opacity, however, a small amount of neutral hydrogen plays a significant role.
The neutral hydrogen beneath the solar convective zone extends down to the point at which $\log(T)\sim 6.7$.
This is the region in which the standard and the enhanced opacities for the solar models  
are different (see Fig. 5).   

The EOS (Rogers, Swenson \& Iglesias 1996) used for OPAL opacity tables, however, is different from the MHD EOS, and based on  
an expansion of physical theories (see, SCVH and Rogers \& Nayfonov 2002). This expansion cannot converge 
in the pressure-ionization regime and consequently reduces the sensitivity of OPAL opacity tables. 


}

\section{Conclusions}
Opacity, as one of the main ingredients of stellar modelling, directly forms inner and outer layers of the stellar interior.
It is effective in the nuclear core and stellar envelope because it controls energy transport. Therefore, any deviation
between theoretical and observational results may arise from a possible uncertainty in opacity.
{ Considering} opacity enhancement as a function of temperature does not solve the { inconsistencies}
between the models of the Sun and $\alpha$ Cen A and B and their seismic and non-seismic
constraints. However, { opacity enhancement treated as a function of density, ionization degree of 
heavy elements (oxygen),
and temperature (equations 4-5) overcomes}  these problems. 
The enhancement below the convective zone, 
{ 
where pressure ionization is effective, 
}
is about $\delta\kappa/\kappa=0.1$ for the Sun and 0.17 for $\alpha$ Cen B.

{ 
Opacity is { influenced} by the excitation level and ionization degree of ions. 
These processes are function of density and chemical composition as well as temperature.
Therefore, temperature-dependent opacity enhancement may solve problems for a single star but not 
for multiple stars. The physical conditions for a given temperature are different for different stars.
The density at $\log(T)$= 6.5, for example, is 0.56 g~cm$^{-3}$ for $\alpha$ Cen A and 0.65 g~cm$^{-3}$ for $\alpha$ Cen B.
The required opacity enhancement at $\log(T)$= 6.5 for these stars is 7 and 22 per cent, respectively.  
Increase in density at a fixed temperature causes ions to reduce ionization degree. This process enhances the opacity.
}


{ The evolutionary tracks of cool stars are significantly changed by the opacity enhancement (equations 4-5).
The effective temperature of the enhanced model around ZAMS is about 300 K smaller than 
that of the standard model, because luminosity is seriously decreased by the enhancement. 
The effective temperature difference between standard and enhanced models with the same luminosity
is about 100 K. This corresponds to 5 per cent increase in radius. 
A further test of the opacity enhancement as a function of local physical conditions in the radiative 
regions of cool stars
in some binary systems is needed. 
Such an enhancement may also { be an alternative reason for} why variable mixing-length parameter is 
required for the 
late-type stars of the Hyades.
}
\section*{Acknowledgments}
The author thanks W. D\"appen for his critiques and suggestions.
This work is supported by the Scientific and
Technological Research Council of Turkey (T\"UB\.ITAK).

\label{lastpage}


\begin{thebibliography}{99}
\bibitem{yildiz} Asplund M., Grevesse N., Sauval A.J., 2005,  in  Bash
F.N., Barnes T.G., eds., ASP Conf. Ser. 336,
Cosmic Abundances as Records of Stellar Evolution and Nucleosynthesis, Astron.
Soc. Pac., San Francisco, p.25 (AGS2005)
\bibitem{bahc} Bahcall J.N., Serenelli A.M., Pinsonneault M., 2004, ApJ, 614,  464
\bibitem{Basu} Basu S., Antia H. M., 1995, MNRAS, 276, 1402
\bibitem{Basu} Basu S., Antia H. M., 1997, MNRAS, 287, 189
\bibitem{ Christen} { Christensen-Dalsgaard J. 1988, in Advances in Helio- and Asteroseismology, IAU Symposium, No. 123, Ed.
J. Christensen-Dalsgaard and S. Frandsen, p.295 }
\bibitem{Basu} Basu S., Chaplin W. J., Christensen-Dalsgaard J., 1997, MNRAS, 292, 243
\bibitem{Basu} Basu S., Chaplin W. J., Elsworth Y., New R. , Serenelli A. M., Verner, G. A., 2007, ApJ, 655, 660   
\bibitem{Bed} Bedding T.R., Kjeldsen H., Butler R.P., McCarthy C., Marcy G.W., O'Toole S.J., Tinney C.G., Wright J.T. 2004, ApJ, 614, 380   (BK2004)
\bibitem{ Christen} Christensen-Dalsgaard J., Dappen W., Ajukov S. V., Anderson E. R., Antia H. M., Basu S., Baturin V. A., Berthomieu G., Chaboyer B., Chitre S. M. et al., 1996, Sci, 272, 1286
\bibitem{yildiz} Christensen-Dalsgaard J., Di Mauro M.P., Houdek G., Pijpers F., 2009, A\&A, 494, 205, arXiv0811.3989
\bibitem{Clausen} Clausen J.V., Bruntt H., Claret A., Larsen A., Andersen J., Nordstr\"om B., Gimenez A.,  2009, A\&A, 502, 253
\bibitem{Eggenberger} Eggenberger, P., Charbonnel, C., Talon, S., Meynet, G., Maeder, A., Carrier, F., Bourban, G. 2004, A\&A, 417, 235
\bibitem{fggenberger} Feltzing S., Gonzalez G.,  2001, A\&A, 367, 253
\bibitem{Iim}Iglesias C.A., Rogers F.J., 1996, ApJ, 464, 943 
\bibitem{Kjel}Kjeldsen H., Bedding T.R., Butler R.P., Christensen-Dalsgaard J., Kiss L.L., McCarthy C., Marcy G.W., Tinney C.G., Wright J.T.  2005, ApJ, 635, 1281
\bibitem{Kjel}{ Kjeldsen H., Bedding T.R., Christensen-Dalsgaard J., 2008, ApJ, 683, 175}
\bibitem{yildiz} Korzennik S.G., Ulrich R.K., 1989, ApJ, 339, 1144
\bibitem{Miglio} Miglio, A., Montalban, J., 2005, A\&A, 441, 615
\bibitem{mhd4} { Mihalas D., Hummer D.G., Mihalas B.W., \& D\"{a}ppen W. 1990, ApJ, 350, 300 (MHD)}
\bibitem{Mih} Mihalas D., Weibel-Mihalas B., 1999, Foundations of Radiation Hydrodynamics, Dover Publications, New York
\bibitem{Rog}{ Rogers F.J., Swenson F.J., Iglesias C.A., 1996, ApJ, 456, 902 }
\bibitem{Rog}{ Rogers F.J., Nayfonov A., 2002, ApJ, 576, 1064 }
\bibitem{Rox}Roxburgh I.W., Vorontsov S.V. 2003, A\&A, 411, 215
\bibitem{Sox}{ Saumon D., Chabrier G., van Horn H.M., 1995, ApJS, 99, 713 (SCVH)}
\bibitem{yildiz}Simon N. R., 1982, ApJ, 260, 87
\bibitem{yildiz}Soriano M., Vauclair S., 2008, A\&A, 488, 975
\bibitem{yldz} {Y{\i}ld{\i}z M., K{\i}z{\i}lo\u{g}lu N., 1997, A\&A, 362, 187 } 
\bibitem{yldz} Y{\i}ld{\i}z M., Yakut K., Bak\i\c{s} H., Noels A.,  2006, MNRAS, 368, 1941
\bibitem{yldz} {Y{\i}ld{\i}z M., 2007, MNRAS, 374, 1264 } 
\bibitem{yldz} {Y{\i}ld{\i}z M., 2008, MNRAS, 388, 1143 } 
\bibitem{yildiz} Zdravkov T., Pamyatnykh A.A., 2008, CoAst, 157, 385
\end{thebibliography}
\end{document}